\documentclass{TTP_DSL2006}

\usepackage[dvipsone]{graphicx}
\usepackage[intlimits]{amsmath}
\usepackage{amssymb}
\usepackage{exscale}

\begin{document}

\title{ Incommensurate spin-density wave in two-dimensional Hubbard model}

\author{ M.A. Timirgazin \inst{1}$^{\rm{a}}$, A.K. Arzhnikov \inst{1}$^{\rm{b}}$ and A.V. Vedyayev \inst{2}$^{\rm{c}}$}

\institute{Physical-Technical Institute, Ural Branch of Russian Academy of Sciences, Kirov str. 132, Izhevsk 426001, Russia
\and
Faculty of Physics, Moscow State University, Moscow, Leninskie Gory 1, 119991, Russia}
\maketitle

\vspace{-3mm}
\sffamily
\begin{center}
$^{a}$timirgazin@gmail.com, $^{ b}$arzhnikov@otf.pti.udm.ru $^{ c}$vedy@magn.ru
\end{center}

\vspace{2mm} \hspace{-7.7mm} \normalsize \textbf{Keywords:} Hubbard model, spin density wave, charge density wave.\\

\vspace{-2mm} \hspace{-7.7mm}
\rmfamily
\noindent \textbf{Abstract.} We consider the magnetic phase diagram of the two-dimensional Hubbard model on a square lattice. We take into account both spiral and collinear incommensurate magnetic states. The possibility of phase separation of spiral magnetic phases is taken into consideration as well. Our study shows that all the listed phases
appear to be the ground state at certain parameters of the model. Relation of the obtained results to real materials, e.g. Cu-based high-temperature superconductors, is discussed.

\section{Introduction}

\noindent Discovery of the high-temperature Cu-based superconducting compounds has given rise to investigation of the two-dimensional (2D) single-band Hubbard model on a square lattice \cite{Anderson}. Despite its seeming simplicity this model contains the physics which underlies such complex electron effects as ferro- and antiferromagnetism and superconductivity. In the exactly half-filling case the ground state of the model is an antiferromagnetic insulator, which agrees to experimental data for such mother compounds for high-$T_c$ oxides as La$_{2}$CuO$_4$, Nd$_2$CuO$_4$ or YBa$_2$Cu$_3$O$_{6}$. Fairly small compositional changes lead to transformation of magnetic structure and conducting properties of these materials. In the case of La$_{2-x}$Sr$_x$CuO$_4$ which is the most extensively studied system an incommensurate magnetic order is observed by neutron experiments \cite{Matsuda,Fujita}. It is generally accepted that superconducting and magnetic properties of cuprates are closely related. This explains a huge interest to investigation of magnetic state of the Hubbard model aside from half-filling. This task appears to be a rather difficult challenge even in the framework of the mean-field approximation.

Incommensurate magnetic states include spin-spiral (SS) and collinear spin-density wave (CSDW) states which are hardly distinguishable in experiment. In Refs. \cite{Sarker,Timirgazin} SS structure has been shown to be the ground state close to half-filling and in a wide region of the magnetic phase diagram of the 2D Hubbard model. In Ref. \cite{Igoshev} the possibility of phase separation has been studied. The main result was that the coexistence of antiferromagnetic (AF) and SS states is energetically favorable in the vicinity of half-filling. CSDW state is much more difficult for investigation than the SS one, because of need of taking into account higher harmonics and an accompanying charge density wave (CDW). CSDW was shown to be more energetically favorable than ferromagnetic (FM) and AF states \cite{Kato} at certain parameters. But at the moment there is no data in literature concerning relative stability of SS and CSDW states. The only study of a so-called soliton lattice state (a set of AF domains divided by nonmagnetic domain walls which is similar to CSDW) shows that this state is more energetically favorable than the SS state at certain parameters \cite{Ichimura}. The aim of our work is to construct a magnetic phase diagram of the 2D Hubbard model considering all the incommensurate magnetic states: SS states with the possibility of phase separation and CSDW states.

\section{Model and results}

The Hamiltonian of the Hubbard model:

\vspace{-5mm}
\begin{eqnarray}
H=\sum_{\mathbf{j,j'},\sigma}t_{\mathbf{j,j'}}c_{\mathbf{j},\sigma}^{+}c_{\mathbf{j'},\sigma} +U \sum _{\mathbf{j} } n_{\mathbf{j}\uparrow}n_{\mathbf{j}\downarrow} .
\end{eqnarray}

It is convenient for us to write the Hamiltonian through charge and spin densities at site:

\vspace{-5mm}
\begin{eqnarray}
n_{\mathbf{j}\uparrow}n_{\mathbf{j}\downarrow}=\frac14\,n_{\mathbf{j}}^2-{S^z_{\mathbf{j}}}^2.
\end{eqnarray}

We make the mean-field approximation which consists in the neglect of the terms of second order of smallness in the expansion of the charge and spin densities in a mean value and a fluctuation:

\vspace{-5mm}
\begin{eqnarray}
n_\mathbf{j}^2\cong-\bar{n}_\mathbf{j}^2+2\bar{n}_\mathbf{j}n_\mathbf{j}
\end{eqnarray}
\vspace{-5mm}
\begin{eqnarray}
{S^z_\mathbf{j}}^2\cong-{\bar{S^z_\mathbf{j}}}^2+2\bar{S}^z_\mathbf{j}S^z_\mathbf{j}.
\end{eqnarray}

For consideration of CSDW we take a unit cell which size $L$ is multiple of the CSDW period. If the wave vector is directed along the diagonal of lattice $\mathbf{Q}=(Q,Q)$ than a unit cell is chosen elongated in that direction as well. Taking into account the periodicity of magnetic structure, we have $\bar{S}^z_{\mathbf{j},l}=\bar{S}^z_{l}$, $\bar{n}_{\mathbf{j},l}=\bar{n}_{l}$, where $l$ is the site number in the unit cell.

Then the Hamiltonian takes the form:

\vspace{-5mm}
\begin{eqnarray}
H=K+V,
\end{eqnarray}

where the kinetic energy (only nearest neighbor hopping is considered):

\vspace{-5mm}
\begin{eqnarray}
K=t\sum_\mathbf{j,j'}\sum_{l,l'}\sum_{\sigma}c_{\mathbf{j},l,\sigma}^{+}c_{\mathbf{j'},l',\sigma},
\end{eqnarray}

and the interaction:

\vspace{-5mm}
\begin{eqnarray}
V=UN\sum_{l}\left({\bar{S^z_{l}}}^2-\frac{{\bar{n}_{l}}^2}{4}\right)+2U\sum_{\mathbf{j},l}\left(\frac{\bar{n}_{l}}{4}n_{\mathbf{j},l}-\bar{S}^z_{l}S^z_{\mathbf{j},l}\right).
\end{eqnarray}

The Fourier transformations lead to the following form of the Hamiltonian:

\vspace{-5mm}
\begin{eqnarray}
K=t\sum_{\mathbf{k},l,\sigma}\left(c^+_{\mathbf{k},l,\sigma}c_{\mathbf{k},l+1,\sigma}(\mathrm{e}^{ik_x}+\mathrm{e}^{ik_y})+c^+_{\mathbf{k},l,\sigma}c_{\mathbf{k,}l-1,\sigma}(\mathrm{e}^{-ik_x}+\mathrm{e}^{-ik_y})\right)
\end{eqnarray}

\vspace{-5mm}
\begin{eqnarray}
V=UN\sum_{l}\left({\bar{S^z_{l}}}^2-\frac{{\bar{n}_{l}}^2}{4}\right)+U\sum_{\mathbf{k},l}\left[\left(\frac{\bar{n}_{l}}{2}-\bar{S}^z_{l}\right)c^+_{\mathbf{k},l,\uparrow}c_{\mathbf{k},l,\uparrow}+\left(\frac{\bar{n}_{l}}{2}+\bar{S}^z_{l}\right)c^+_{\mathbf{k},l,\downarrow}c_{\mathbf{k},l,\downarrow} \right].
\end{eqnarray}

\begin{figure}[!h]
\hskip 2cm
\includegraphics[width=11cm]{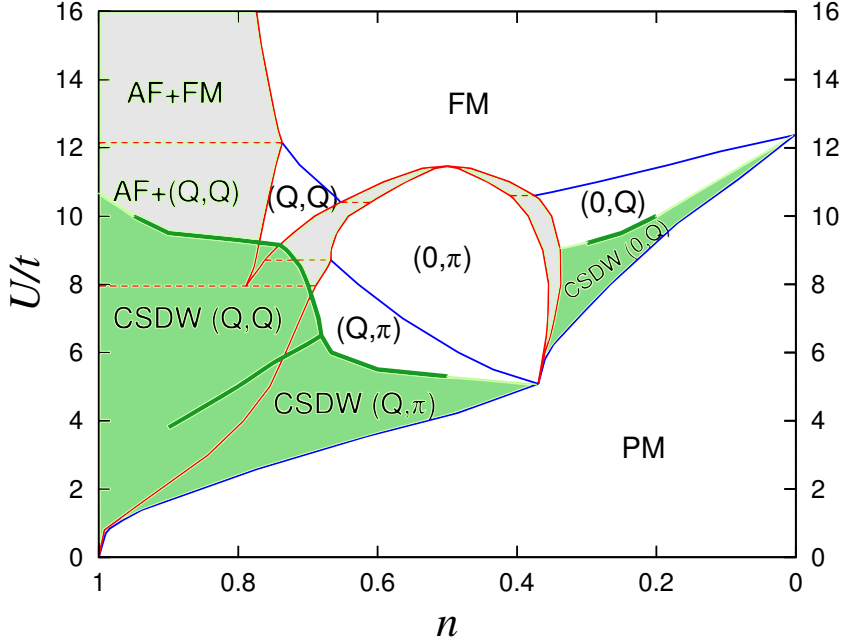}
\caption{(Color online) Magnetic phase diagram of the 2D Hubbard model on a square lattice. Bold green lines denote calculated boundaries of CSDW state, blue lines denote phase transitions between homogeneous SS states, red lines denote phase transitions between separated SS states. Regions of CSDW state are dark shaded (green), regions of phase separation are light shaded (grey). } \label{diagram}
\end{figure}

\begin{figure}[!h]
\hskip 2cm
\includegraphics[width=11cm]{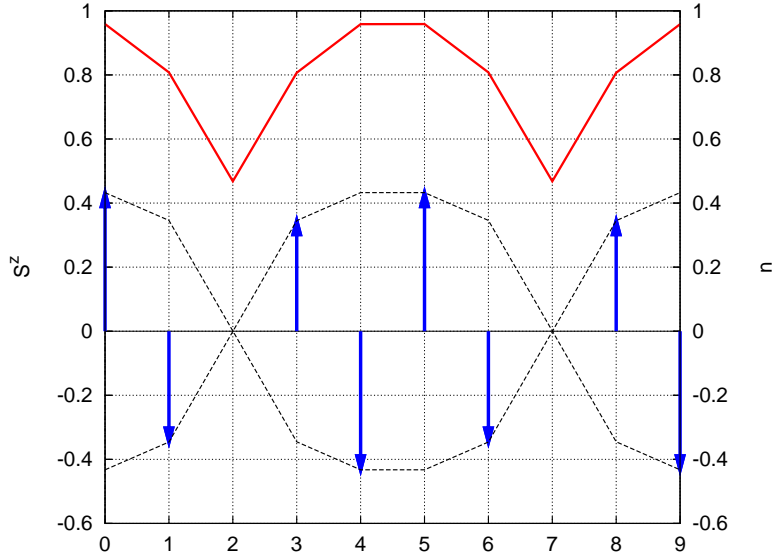}
\caption{The spatial profiles of spin (lower plot) and charge (upper plot) densities as a function of lattice sites for $U=8$, $n=0.8$.} \label{sdw}
\end{figure}

Similarly one can obtain the Hamiltonian for parallel states $(Q,\pi)$ and $(0,Q)$. There considered are the diagonal and parallel CSDW states only because of their relative simplicity and on the basis of data obtained for SS states for which only diagonal and parallel states were shown to be the ground ones \cite{Timirgazin}.
To diagonalize this Hamiltonian we should diagonalize $2L\times 2L$ matrices. The total energy of the system can be find by solving the self-consistent system of equations on charge and spin densities at sites. We suppose that these densities can take any values through the self-consistency procedure, i.e. we do not fix a profile of CSDW and CDW. For fixed $U/t$ and $n=\frac{1}{L}\sum_\mathbf{l}\bar{n}_\mathbf{l}$ we go over different wave vectors $\mathbf{Q}$ and search one that provides a minimum of the total energy. We compare this minimal value with the energy of the ground SS state with phase separations which is calculated by the numerical scheme described in Ref. \cite{Igoshev}. Thus we find the overall ground state considering different incommensurate magnetic phases. The results of calculations are presented in the phase diagram (Fig. \ref{diagram}). We see that the CSDW states appear to be more stable than the SS ones for small and intermediate $U/t$ both in the vicinity of half-filling and far from it. It should be noted that the wave vector which provides the minimum of the energy can be, as a rule, calculated by formula $Q=\pi n$ (for $(Q,Q)$ and $(Q,\pi)$ states). The boundary of CSDW and PM states is supposed to coincide with the boundary of SS and PM states --- accuracy of calculations does not allow to determine this line more precisely.  On Fig. \ref{sdw} spatial profiles of CSDW and CDW are presented for $U/t=8, n=0.8, \mathbf{Q}=(0.8\pi,0.8\pi)$. One can see a strong deviation from the harmonic wave form. The larger $U/t$ the closer magnetic structure to a soliton lattice state, i.e. antiferromagnetic domains separated by nonmagnetic domain walls. As follows from the figure, CDW has sufficiently large amplitude --- from 0.47 at nonmagnetic sites to 0.96 on sites with maximal magnetization. Such a strong charge inhomogeneity looks rather unrealistic. An additional Coulomb energy caused by this inhomogeneity is not taken into account in the framework of our model, but it is rather obvious that it should substantially raise the energy of the CSDW so that it would not be the ground state.

\section{Summary}

\noindent We have constructed the magnetic phase diagram of the 2D Hubbard model in the mean-field approximation. The calculations show that the CSDW state is the ground state in a wide region of the model parameters. It is more stable than the uniform SS state and the SS state taking into account phase separations. We did not take into consideration an additional Coulomb interaction rising from the charge inhomogeneity of the system. We believe that this interaction could be sufficiently strong to suppress CSDW state. That is why we think that the true nature of the CSDW state stabilizing in real systems is beyond the mean-field treatment and is still open to question.

\section{Acknowledgments}

\noindent The work is supported by grants No. 09-02-00461 and No. 10-02-90708 from Russian Basic Research Foundation, and by Program of RAS No. 09-2-2001.

\end{document}